# Orchid: Orchestrating Context Across Creative Workflows with Generative AI

SRISHTI PALANI*, Tableau Research, USA
GONZALO RAMOS, Microsoft Research, USA

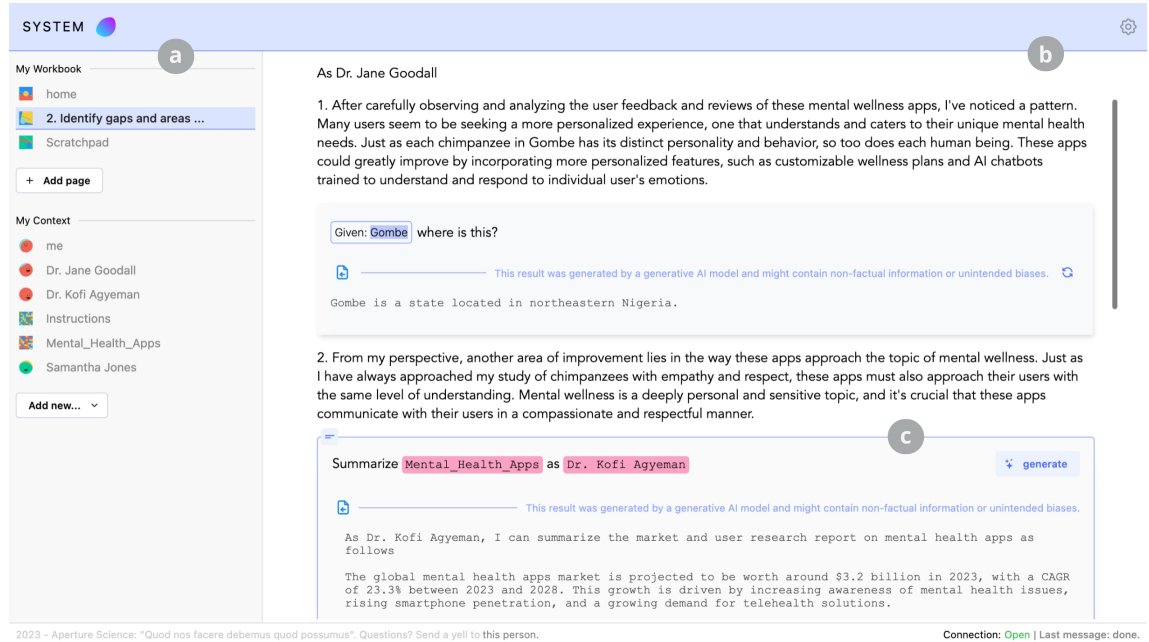

Fig. 1. *Orchid*'s UI. a) *Documents Sidebar*. b) *Editor Area*. c) a result block from a *Summary* operation referencing context in a market research document using a persona's style.

Context is critical for *meaningful* interactions between people and Generative AI (GenAI). Yet mainstream tools offer limited means to orchestrate it, particularly across workflows that span multiple interactions, sessions, and models, as often occurs in creative projects. Re-specifying prior details, juggling diverse artifacts, and dealing with context drift overwhelm users, obscure intent, and curtail creativity. To address these challenges, we present *Orchid*, a system that gives its users affordances to specify, reference, and monitor context throughout evolving workflows. Specifically, *Orchid* enables users to (1) *specify* context related to the project, themselves, and different styles, (2) *reference* these via explicit mentions, in-line selection, or implicit grounding, and (3) *monitor* context assigned to different interactions across the workflow. In a within-subjects study (n=12), participants using *Orchid* to execute creative tasks (compared to a baseline toolkit of web search, LLM-based chat, and digital notebooks) produced more novel and feasible outcomes, reporting greater alignment between their intent and the AI's responses, higher perceived control, and increased transparency. By prioritizing context orchestration, *Orchid* offers an actionable step toward next-generation GenAI tools that support complex, iterative workflows — enabling creators and AI to stay aligned and augment their creative potential.

## 1 INTRODUCTION

Context plays a critical role in how we interact with the world, shaping how information is communicated, understood, and applied. In interactions between humans and computers, especially with Generative AI (GenAI), context enables alignment between user intent and system responses. It helps interpret individual utterances: resolving ambiguity and ensuring continuity in anaphoric and cross-referential phrasing (e.g., [29, 32, 45]). Beyond single interactions, context becomes even more vital when maintaining a coherent thread across complex workflows that span multiple interactions, sessions, and iterations [15, 37, 40]. Context also underpins collaborative workflows, where shared understanding of context among agents – be they humans or AI – becomes the foundation of effective collaboration [14, 33, 41, 42]. In essence, human-AI teams must remain "on the same page" by continually establishing a shared understanding of what is relevant at any given point.

Creative workflows exemplify the complexity of and need for orchestrating context: These workflows are inherently iterative, nonlinear, and span multiple tasks and sessions [1, 15, 43]. Collaboration is often integral, involving either human collaborators or GenAI models acting as creative partners [39, 40]. Previous research has highlighted the importance of maintaining a cohesive "thread" across individual interactions, a challenge amplified by the diversity of tasks, modalities, and iterations involved in creative workflows [14, 37, 42].

While recent advances in GenAI and Creativity Support Tools (CSTs) show promise, they fall short in facilitating seamless context orchestration. On the GenAI front, tools such as ChatGPT, Claude, and Gemini now offer features like longer context windows (e.g., up to 100K tokens [3, 11]), canvas-based interfaces[1], and project workspaces[2]. These enhancements enable larger swaths of text or media to be considered all at once, potentially retaining more information about a user's ongoing task. However, simply increasing context window size is not a panacea for complex workflows: when the context becomes too large or dense, it can lead to context drift – where the model either ignores important details or produces irrelevant, off-track responses due to difficulty identifying which portions of the context are most relevant [9, 30]. On the CST front, popular tools have begun integrating GenAI, like Figma's AI plugins[3], Notion AI[4], and Google's Duet AI[5], where based on a user prompt, GenAI can suggest new text and graphic elements, summarize, and improve existing content. While such features help quickly generate content, they are not designed to synthesize and maintain context across individual interactions. This forces creatives to repeatedly re-specify context or manually piece together details across multiple interactions in the CST [37, 40, 44]. This continual context shifting by creatives using GenAI tools can lead to cognitive overload and hamper creative potential, emphasizing the critical need for solutions that *intelligently integrate and manage context* – including the ability to constantly establish a shared understanding between the user and AI systems.

To address these gaps, we present *Orchid*, a set of interaction mechanisms designed to orchestrate context across creative workflows. *Orchid* enables users to (1) specify and externalize context in the following ways: *(i) Project context:* Users can organize and reference projects' working context within *workbook pages*. Supporting materials, (e.g.

---

[1]ChatGPT Canvas: https://openai.com/index/introducing-canvas/ and Claude Artifacts: https://www.anthropic.com/news/artifacts
[2]Claude Projects: https://www.anthropic.com/news/projects ChatGPT Projects: https://help.openai.com/en/articles/10169521-using-projects-in-chatgpt
[3]Figma's AI plugins: https://www.figma.com/ai/
[4]Notion AI: https://www.notion.com/help/guides/category/ai
[5]Google Duet AI: https://workspace.google.com/blog/product-announcements/duet-ai

collaborators' research, meeting notes, data, etc.) that are relevant to the project, can be *uploaded* to enrich the GenAI's understanding. Context from project planning and task management is specified in the *home page*. *(ii) User's Personal context:* Inspired by reflective journaling practices, users can share their emotional state, design preferences, or other personal insights via a *Me page*. *(iii) Stylistic context through personas:* Users can simulate collaborators or experts by creating and managing "personas" and their skill sets, characteristic styles, personality traits, etc., in *Persona pages*. More importantly, *Orchid* provides affordances to (2) reference relevant context – both implicitly and explicitly – and crafts *'meta-prompts'* that seamlessly integrate context into GenAI interactions. For example, users can: (i) *Explicitly reference* relevant context using "@" shorthand, (ii) use *In-line selection and prompt* to specify the relevant context for a prompt, (iii) *Implicitly ground interactions* in a page's content or associated persona simply by prompting within that specific type or part of the age. So even if a user's prompt is brief, fuzzy or ambiguous, *Orchid* leverages these mechanisms to ensure the output remains context-aware and aligned with user intent, helping maintain a shared understanding between user and AI. (3) To monitor and manage which context was assigned, implicitly or explicitly, to each interaction, hovering over the generate button or result block of GenAI interaction activates the Transparency Lens.

We evaluated *Orchid*'s interaction mechanisms' potential to orchestrate context across a workflow by conducting a within-subjects user study ($n = 12$). Participants were asked to engage in creative projects using both *Orchid* and a baseline condition consisting of their usual set of tools such as web search, LLM-based chat, and a digital notebook. Our findings reveal that: Participants generated a greater number of ideas that were more novel, feasible, and creative. Orchid's key benefits revolve around reducing repetitive prompting by letting users explicitly and implicitly reference project documents, personal preferences, and simulated expert "personas", which in turn increases alignment between human intent and AI outputs. Participants also felt more control, transparency, and emotional support, describing Orchid as more of a collaborative partner than a mere tool. Overall, the study highlights that supporting context orchestration can enhance both creative outcomes and the users' perception of AI as an empathetic teammate, rather than just a prompt-response engine.

This paper contributes:

(1) *Orchid* a tool that illustrates interaction techniques for specifying, referencing, and monitoring different types of context – including project details, user preferences, and stylistic personas – across creative workflows.
(2) Empirical insights from a within-subjects user study (*n*=12) comparing *Orchid* to a baseline of fragmented tools: expert raters found its outputs more novel, feasible, and valuable. Participants also felt better alignment with their goals, greater control, and a more collaborative relationship with GenAI.

## 2 RELATED WORK

### 2.1 Establishing Common Ground: Role of Context in Human-to-Human Interaction

Humans excel at communicating effectively because they go beyond the literal meaning of words to infer the speaker's intentions and underlying meaning. One of the core processes underpinning this is grounding, or the ongoing process of establishing and maintaining mutual understanding in conversation [8]. For humans, grounding occurs intuitively through shared knowledge and incremental confirmation. However, for systems, grounding relies on explicit mechanisms for context management, such as tracking conversational history or interpreting cross-references [17, 36].

Context plays a crucial role in providing the implicit, situational, and relational cues for grounding. Prior work in Human-Computer Interaction (HCI), Linguistics, and Natural Language Processing (NLP) have identified several distinct types of context that shape interaction, including: (1) *Linguistic Context:* Details such as discourse structure, anaphora, and reference resolution [8, 17]; (2) *User Context:* User goals, intentions, emotional states, and preferences [13]; (3) *Social/Collaborative Context:* Awareness of other collaborators' actions, roles, knowledge and styles [18, 41]; (4) *Task Context:* Information about the task structure and constraints that influence how people plan and carry out work [38]; (5) *Situational/Environmental Context:* Physical location, device capabilities, and external constraints [12]. Each of these context types can influence how systems interpret user utterances, what responses are appropriate, and how collaboration between humans or humans and systems unfolds. Understanding and leveraging these nuances helps ensure that systems align with user intent, mitigates ambiguity, and supports continuity across interactions.

Our work takes heed on the importance of context and provides affordances to specify and reference many of these types of context as first-class actions in the user experience of *Orchid*.

> **Design Goal 1**
>
> To support continuous shared understanding between users and GenAI, systems should enable the specification, persistence, and monitoring of diverse context types—including task, personal, and stylistic context—so that AI outputs remain aligned with evolving user intent.

### 2.2 Addressing Context Challenges in Human-AI Interactions

While large language models (LLMs) enable richer interactions through natural language, they struggle to fully capture and incorporate context. In NLP, context encompasses the surrounding text or speech that influences interpretation. This includes words, phrases, sentences, or larger discourse structures that precede or follow a linguistic unit [22]. For example, the meaning of "bank" depends on whether it refers to a financial institution or the side of a river, determined by its linguistic and situational context. Advances like GPT-4o1, Claude and Gemini feature extended context windows to accommodate larger inputs [3, 5, 11]. However, simply lengthening the window introduces new problems such as context drift, where the model cannot reliably identify or weigh the most relevant information [9, 30].

To mitigate these limitations, researchers have explored techniques such as retrieval-augmented generation (RAG), which involves indexing external text corpora – e.g., knowledge graphs – and fetching relevant snippets in real time [19, 20, 28]. This allows the LLM to consult a broader knowledge pool while staying focused on information deemed relevant to the user's query. Additionally, some systems employ long-term memory modules or hierarchical prompting techniques to organize context at multiple levels of granularity [6, 23]. Although promising, these strategies often require sophisticated implementations and do not inherently address how the *user* wants to specify, track, and recall context across multi-model and multi-session workflows [37]. This gap underlines the importance of designing systems like *Orchid*, which aim to bridge these challenges by enabling users to explicitly and implicitly specify, reference, and orchestrate context across multiple interactions, sessions, and models interacted with over the course of a workflow.

Our work reflects on these challenges and builds on them by introducing interaction mechanisms in *Orchid* that allow users to explicitly and implicitly reference relevant context. On the back end, *Orchid* generates a more contextual meta-prompt, using appropriate prompt engineering techniques to manage and stitch together contextual details – designed to produce outputs that more aligned with the user's workflow.

> **Design Goal 2**
>
> To reduce redundancy and improve relevance, systems should provide affordances for users to implicitly and explicitly reference specific context across interactions – and synthesize that context into structured, model-readable prompts.

### 2.3 Contextual "Threads" in Creative Workflows and GenAI-Enhanced CSTs

Creative workflows are inherently iterative and nonlinear, spanning multiple tasks, sessions, and collaborators [10, 43]. In such workflows, context acts as the thread that connects ideas, decisions, and actions over time, ensuring alignment with the project's goals and vision. For instance, a designer collaborating with an AI tool to create a website requires context such as the target audience, branding guidelines, design principles, and feedback from collaborators. Without this context, the AI might generate visually appealing designs that fail to meet functional or stylistic requirements.

Prior research has emphasized the critical functions of context in creative workflows [14, 37, 40, 42, 44]. Context anchors intent by providing a shared understanding of the project's objectives, constraints, and results, ensuring that efforts remain aligned. It preserves continuity across iterations by retaining prior work, feedback, and intermediate ideas, reducing the risk of losing critical information during refinement. Additionally, context lowers cognitive load by externalizing key details, allowing creators to focus on ideation and execution rather than reconstructing prior decisions. In collaborative settings, shared context acts as common ground, enabling effective communication between collaborators – whether human or AI – by clarifying roles, contributions, and how they integrate into the overall workflow. Furthermore, context stimulates creativity by fostering associations between ideas, facilitating exploration of innovative combinations.

More recently, Creativity Support Tools (CSTs), like Figma, Miro, and Notion, have recently integrated generative AI to expedite tasks such as brainstorming, prototyping, and content generation [40]. So have emerging tools in HCI research, like SpellBurst [2], PersonaFlow [31] AINeedsPlanner [24]. However, most of these features treat context superficially – often only referencing a single document or prompt at a time [44]. Because creative work typically unfolds across multiple iterations and collaborators, poorly managed context can result in repetitive prompting, missed references, and confusion over versioning and rationales [37, 42]. Furthermore, collaborative creative endeavors rely on shared mental models and dynamic role assignments, yet current solutions rarely ensure continuous context alignment throughout the workflow between the human and the participating AI. Building on these insights, the *Orchid* system introduces a new approach that prioritizes *user-driven context specification* alongside methods to reference, ground, and orchestrate context across the creative workflow—addressing a significant gap in existing GenAI-based CSTs.

Building on prior work emphasizing the importance of 'contextual threads' in creative workflows, our approach embraces the non-linear nature of this work through a flexible notebook metaphor – where pages do not impose predetermined structures – and provides an integrated environment to prevent context from fragmenting across multiple applications.

> **Design Goal 3**
>
> To support iterative and collaborative creative workflows, systems should enable users to structure and traverse contextual threads—spanning goals, tasks, artifacts, and collaborators—so both humans and GenAI can build on prior work coherently and without rework.

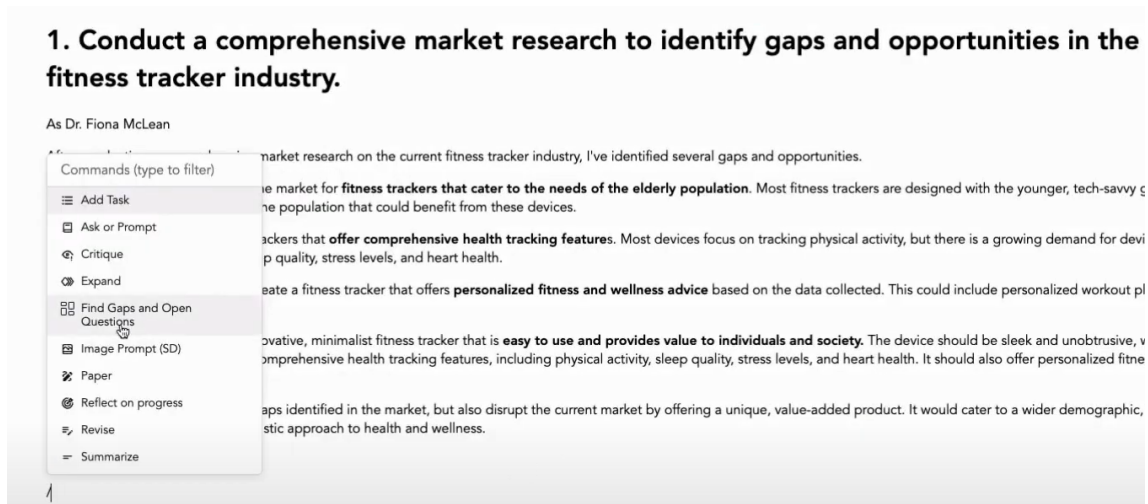

Fig. 2. "/" (Forward Slash) Commands: Users can access various creative operations by typing "/" to open the drop-down menu and selecting the desired operation.

## 3 ORCHID SYSTEM

To provide users with a familiar user interface to work with and reduce the effort of discovering features and how to use them, *Orchid*'s base user interface is a workbook modeled after popular digital note-taking interfaces that most users would have encountered, like Notion and Microsoft Loop [6]. *Orchid*'s front-end interface comprises two main areas: the *Documents Sidebar* and the *Editor Area* (Figure 1). The *Documents Sidebar* (Figure 1) is a component that allows users to select, create, and delete documents. When a particular document is selected, it appears in the *Editor Area*, which takes most of the area of the interface and is the area where people can view and edit the currently selected document.

Similar to other GenAI features in CSTs, on every page in *Orchid*, users can invoke GenAI models with the *"/" command* to invoke a drop-down menu that allows them to select which (creative) operation they would like to access. Currently, the list of creative operations a user can ask GenAI to do includes generation based on prompt, searching the Internet, critiquing, expanding on content, finding gaps and open questions, reflecting on progress towards project goal, revising, and summarizing. All *task* and *operation* components have a basic structure comprising a text entry field and an action button that triggers a particular type of operation. Upon hovering over the action button, they can select the level of unconventional generation they want: precise, balanced, or creative (Figure 4).

### 3.1 Specifying Context

*Orchid* enables users to specify and externalize different types of context **(DG1)**:

- **Project context:** Information about the project and how work is being carried out. There are three ways in which creatives can specify project context:
  - **Workbook Pages**: These are regular rich text documents that the user can edit in the same way that they edit a rich text document on an online editor. These show the work done by the user, e.g., sketches, drafts, and other intermediate artifacts (see '2. Identify gaps and areas ...' page in Figure 1's Documents Sidebar).

---
[6]https://loop.microsoft.com

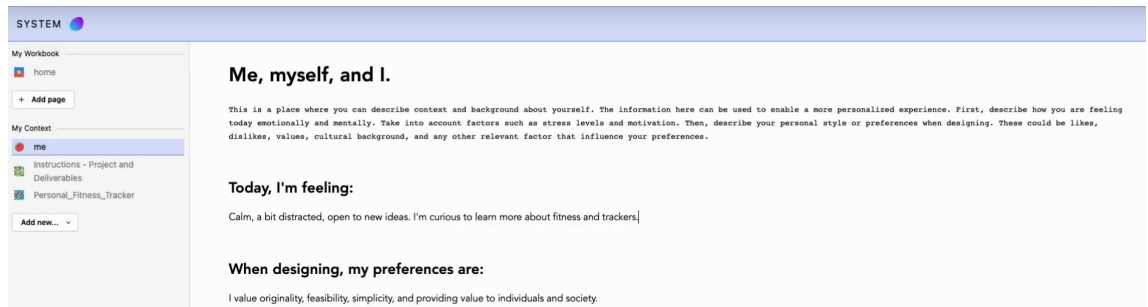

Fig. 3. Me page: Users can edit this text file to define how they would like to be seen by the system externalizing personal context such as their current emotional state and design preferences

- **Context Pages** These are uploaded documents, external to the project but still relevant. For example, the market research document shared by their teammates, or client requirements document (i.e., 'Mental_Health_Apps' and 'Instructions' pages under 'My Context' in Figure 1's Documents Sidebar).
- **home page**, where users can plan the project, create and assign tasks. Here, users can enter their creative goals into the *goal component*, and clicking on 'execute goal' will break down the goal into specific, achievable, and relevant tasks and display them as *task components* (Figure 4). Users can specify how certain the action plans generated are by adjusting the temperature of the goal component. When a user starts a task, a dedicated working page is created for it, and subsequent generations about this task will appear on it.
- **User Context: me page** in the My Context section, which is the place where one can define how they would like GenAI to see them by externalizing personal context such as their current emotional state, working style, and creative preferences.
- **Stylistic Context:** Users can create and channel perspectives of custom **'Persona pages'** by creating a new persona under 'My Context' in *Orchid*'s sidebar (see Dr. Jane Goodall, Dr. Kofi Agyeman, and Samantha Jones in Figure 1). This technique enables users to modulate GenAI model behavior by creating and managing 'expert personas' and their skill sets, characteristic styles, personality traits, etc., outside of prompts (Figure 5). If they are not sure what might be interesting or relevant perspectives to consider, the user can ask the system to generate relevant personas for each task by clicking the *'Generate persona'* button in each 'Task' component on the 'Home' page. This adds a persona page in the *Documents Sidebar* context area. If one is not happy with the persona generated, one can click the action button again and generate a new one, or alternatively, edit the persona's definition in the context area. From then on, if the user starts that task, it will be carried out from the lens of the persona associated with it. This is inspired by an insight from [37], where creatives wanted interactions with GenAI to model social relationships in creative studios, such as design collaborations and critique sessions. Also, inspired by role-play prompting [26] a popular and effective prompt engineering technique.

### 3.2 Refering to Context

We designed *Orchid* to enable users to better align LLM outputs with their intentions by explicitly or implicitly referring to relevant contexts present in the many artifacts generated across their workflow **(DG2)**. These interaction techniques aim to help maintain a thread across outputs generated over the course of a workflow without repeating relevant context and improves the reliability of the generated responses. Users can ground any generation in a relevant context using one of three ways:

(1) **Explicitly Refering:** By typing in '@', another familiar notebook interaction pattern, and then selecting the relevant context from the drop-down list of project-relevant context (see Figure 6).
(2) **Implicitly Grounding:** by placing the creative operation or prompt next to the relevant context that is somewhere in the notebook. This takes in the working page it is placed on, and any references on this page as the relevant context (see Figure 8 where the prompt is short, but its placement takes the page's content and persona as context).

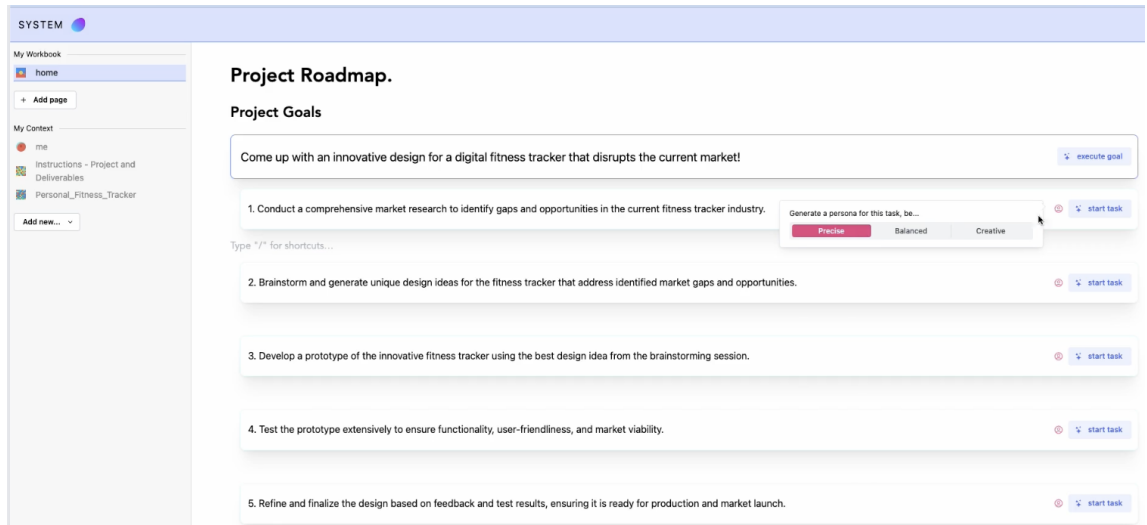

Fig. 4. Users can provide a high-level objective, which the system can then break down into actionable tasks. Also, all following GenAI outputs will use this as additional context

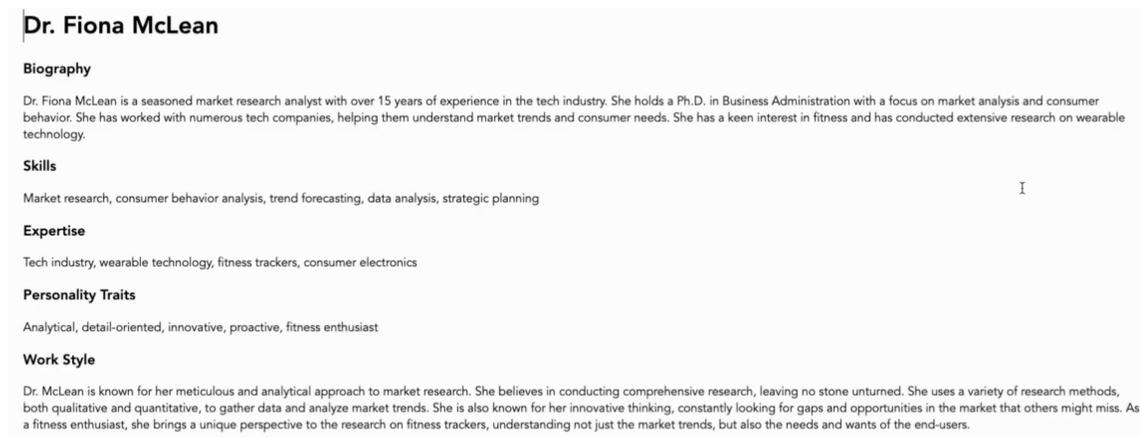

Fig. 5. Example of a persona page. Users can edit this text file to customize their personas to best help with their work.

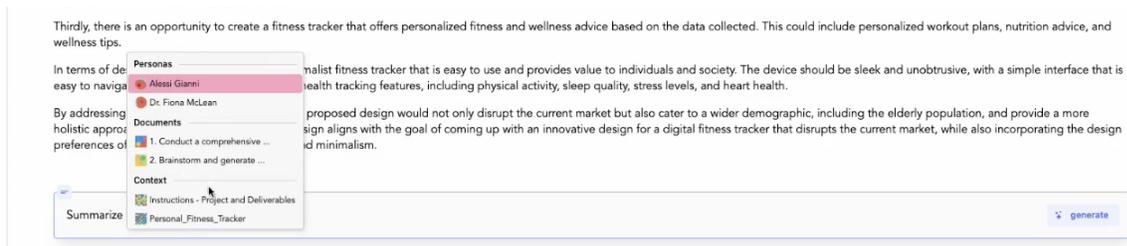

Fig. 6. This Summarize component is an example of a "/" operation. Here you can see how the user can make use of '@' mentions to change how the operation is grounded to different personas and/or documents.

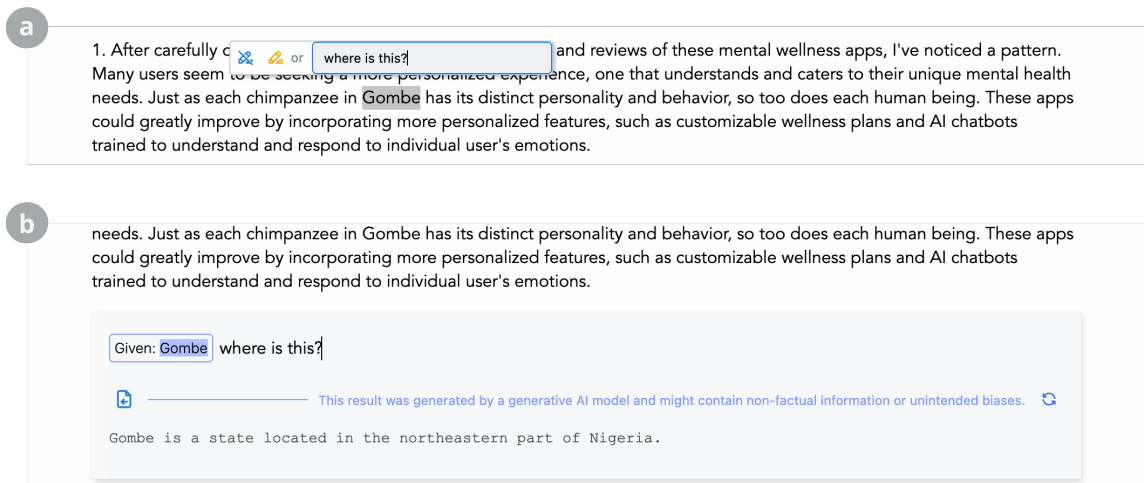

Fig. 7. In-line prompting. This figure illustrates *Orchid*'s support for in-line contextual prompting. a) first the user will select a part of a document, which will reveal a contextual prompt area where the user enters their request. b) after the request is fulfilled, it is inserted as a block, as the next paragraph after the selection. As with any results block, the user can regenerate the result, change the generation parameters, paste the result into the page, or delete the block.

(3) **In-Line Prompting:** User can select relevant content the same way as one would do in any text editor (e.g., clicking and dragging the cursor). This brings up *Orchid* In-Line Prompt that allows users to apply a prompt to the current selection. After entering the prompt, a context prompt block is inserted immediately after the closest paragraph containing the selection (see Figure 7). This lightweight capability is not grounded to a page or associated with a persona unless otherwise specified.

All "/" operations have a default grounding and persona. The default grounding is the page that contains the prompt component. If the page was created by a task prompt, it will be associated with that persona, and all prompts in it will be associated with that persona by default. If no persona is associated with a page, then there is no persona associated with a prompt, and their results are produced through the lens of a generic "AI assistant."

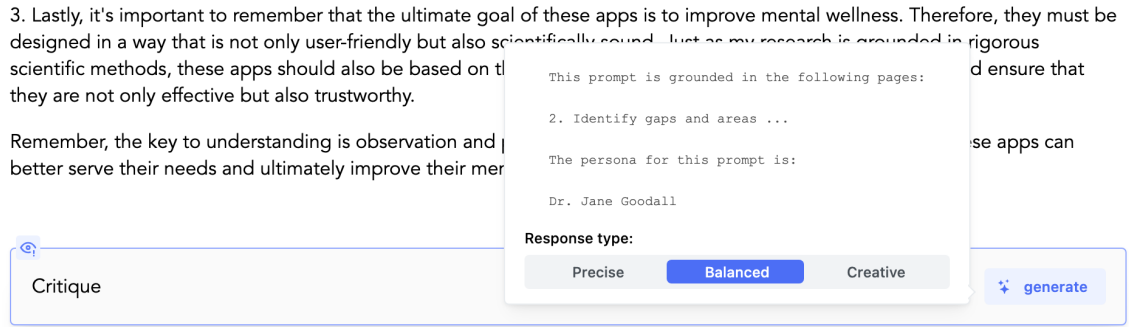

Fig. 8. Transparency lens: Hovering over the generate button of each "/" operation component displays what context it is grounded on so that the users know what information it has access to for transparency and in case they want to change it.

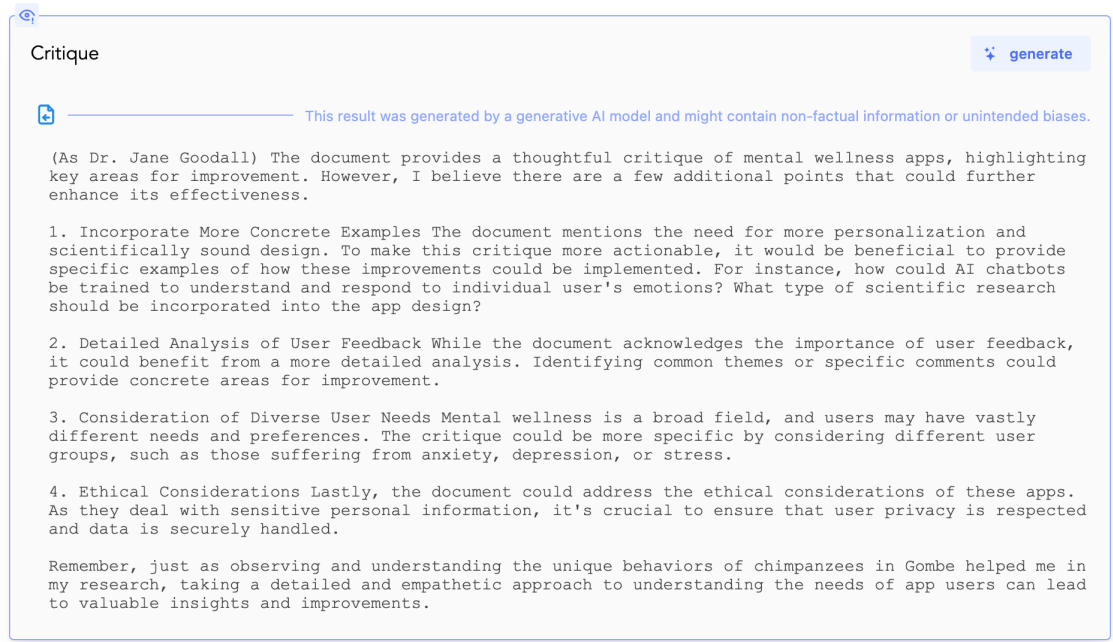

Fig. 9. Critique component results. The figure illustrates how the result of a "/" operation is presented. The *block* keeps the results contained and give users the option to regenerate the results, insert them into the document, or deleting it altogether.

### 3.3 Monitoring Context

To monitor and manage context for each GenAI interaction, *Orchid* has **Transparency Lens (DG1 & DG2)**. Hovering over the generate button of each creative operation or prompt component displays what context it is grounded on (i.e., which page, persona, etc.) so that the users know what information it has access to for transparency and in case they want to change it.

After the user submits any creative operation or operation prompt, *Orchid* presents the results of the operation in a **Result block** that allows users to preview, regenerate, discard, or insert the result in the containing page **(DG2 & DG3)** (Figure 9). *Orchid*'s support depends on asynchronous operations, that is, performing a GenAI-based operation that does not produce immediate results (e.g., a GenAI API service throttling). *Orchid* embraces these delays and has its prompt components and result blocks reflecting their "in progress" status in a way that does not prevent the user from performing other activities (e.g., switching pages, further editing, prompting) on the same page or another page while results are on their way.

### 3.4 Implementation Details

The *Editor Area* is built on top of TipTap[7], a headless open-source editor, which we enhance with custom components to invoke and present bespoke AI capabilities.

*Orchid*'s functionality and capabilities are supported by a *Semantic* and *Document Services* back-end that exposes a RESTful endpoint for AI capabilities and document management. Document services consist of providing an endpoint for CRUD operations to manage and maintain the different types of documents supported by *Orchid*. This role allows for document persistence beyond a browser session and for access to documents' contents when semantic operations are performed.

The back end provides AI semantic capabilities by building on top of general LLM functionality. *Orchid*'s back-end uses LangChain [8] and OpenAI's GPT-4 [35] to fulfill a user's request for a particular capability or operation. Figure 10 illustrates the core structure in which a system's operation is fulfilled: a user's prompt is expanded using grounding and contextual information into a meta-prompt that is passed to an LLM, in turn the response is (parsed) and passed back to the user. For details about the prompts used in *Orchid*, readers should refer to the Supplementary Materials **(DG1 and DG2)**.

In the back end, each prompt is augmented with relevant information by leveraging prompt engineering techniques like chain-of-thought-reasoning [46], Retrieval-Augmented Generation [16] and ReACT [47] agents with access to tools including a search engine, other GenAI models such as stable diffusion, etc.

## 4 USER EVALUATION STUDY

To investigate *Orchid*'s effectiveness in orchestrating context across creative workflows, we conducted a within-subjects study (n=12). We asked participants to complete a complex creative task using *Orchid* as well as a baseline condition consisting of off-the-shelf tools generally used in creative workflows such as a search engine, an LLM-based chat tool, and a digital notebook.

### 4.1 Tasks

We presented the project to participants as a simulated work task scenario [4]:

> "Imagine you are a designer working for a client to redesign [ Airport Security Experience || Fitness Trackers || Personal Finance Management Apps || Job Search Apps || Mental Health Apps]. Your job is to change the existing design concept so that any identified issues are solved and the product is more innovative, original, and/or valuable.
> During the task, you will create:

---
[7] https://tiptap.dev
[8] https://www.langchain.com/

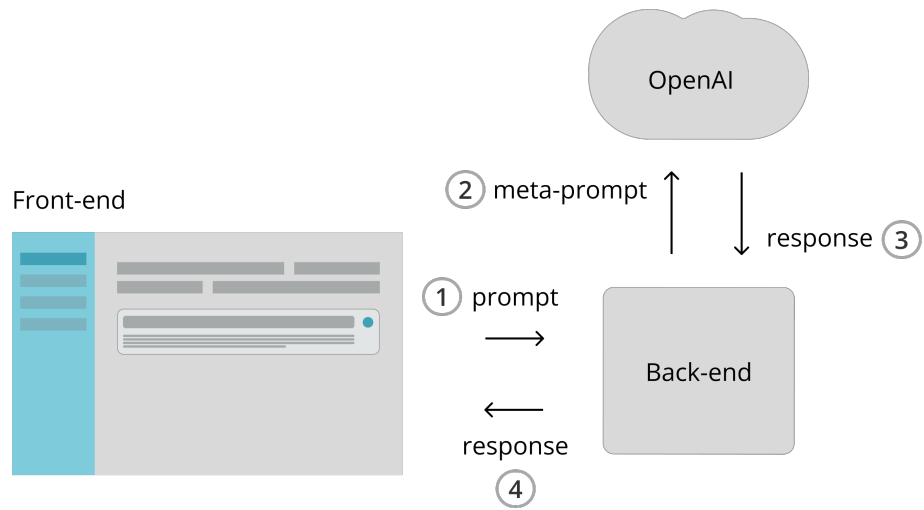

Fig. 10. *Orchid*'s components. This figure illustrates at a high level how the front end and the back end interact. 1) when an operation takes place, the prompt visible in the component is sent to the back-end. 2) the back-end enhances the prompt into a meta-prompt using its knowledge about the user and operation context; the meta-prompt is then sent to the LLM. 3) the LLM's response if received by the back-end, which can perform parsing and formatting operations. 4) the back-end sends the parsed response to the front-end, which can further parse it, then presents it into a response block.

- 2-3 original ideas of how to make the product better
- 1 final refined idea
- Script for a short elevator pitch

As part of the within-subjects study, each participant completed the creative task once with Orchid and once with the baseline, using two different topics to avoid learning carryover. As a potential starting point, they received a market and user research report to understand the product, its market, and user needs. The study lasted a total of 2.5 hours, each task for 45 minutes, with a 10-minute break in between. Topics were complex but familiar enough for participants to build on prior knowledge while exploring new ideas.

### 4.2 Baseline

To choose an ecologically-valid baseline, we compared *Orchid* to the tools that are representative of how users currently perform creative processes. Although there were no limitations on what people could use during the baseline condition, we recommended participants to stay within the tools from the following list: Internet browser (e.g., Google Chrome, Safari, or Microsoft Edge), LLM (e.g., Bing Chat, ChatGPT, Claude), and Search engine (e.g., Bing, DuckDuckGo, Google). We also provided a text editor as a place to take notes, paste content, and process information (e.g., Notion or Microsoft Loop). This list of tools is derived based on insights from our formative study in which we asked participants what tools they used during their creative process.

### 4.3 Participants

We recruited 12 participants (four female, eight male) across research, design, and software engineering departments of a large technology company via mailing lists. In terms of their roles at work, one was a writer, one was a designer, four were engineers, two were people managers, two were doctoral students, and two were researchers. Eight of them had accumulated 6-10 years of experience in their respective fields, while two possessed less than a year of experience, and another two had between three and five years of expertise. In terms of experience using generative AI models, two said less than 6 months, six of them reported having experience of 6 months to 1 year, two for 1-2 years, and one for 2-3 years. Seven participants reported using these models very frequently (multiple times a week), four reported using them frequently (multiple times per month), and one reported using them occasionally (a few times a year). We conducted all studies remotely over video calls. We compensated each participant with a $100 USD gift card for an approximately 2.5-hour study.

### 4.4 Study Procedure

Before the study appointment, participants were sent an informed consent form and asked to complete a demographic survey. During their study appointment, each participant underwent two creative tasks in which the presentation order for the experience condition (*Orchid* and *Baseline*) was counterbalanced.

During the 45-minute task, participants freely interacted with the tool(s) in each condition to generate between two and three original ideas of how to make a product better, converge on a final refined idea, and write a short elevator pitch, all while thinking aloud about their experiences [21, 27]. Before starting each 45-minute task, participants either watched a 5-minute tutorial for the *Orchid* condition or were given a short refresh on the *Baseline* set of tools. We gave them approximately 15 minutes to explore and practice.

After each condition, we asked participants to rate their level of agreement with statements about their user experience. We also asked them to list up to three things they liked, as well as up to three things they would like to improve about their experience with each condition. At the end of the study, once they had experienced both conditions (*Baseline* and *Orchid*), we asked them which one they preferred using and why. From the study, we capture self-report data through think-aloud and responses to surveys, as well as application log data, to quantitatively measure and understand user behavior.

### 4.5 Measures

*4.5.1 **Quality of Creative Outcomes**.* User-experience design experts (i.e., designers with at least a Master's degree in Design and completed at least 3 design projects), blind-to-condition rated the three types of creative outcomes (2-3 original ideas of how to make the product better, 1 final refined idea, and Script for a short elevator pitch) on a scale of 1-5 (where higher is better) along the following criteria [43]:

- *Novelty*: refers to the degree to which an idea, product, or solution is unique or original. It is the opposite of something that is obvious, ordinary, or already well-known.
- *Feasibility*: refers to the practicality or workability of an idea or solution. It considers factors such as available resources, technical constraints, economic viability, and compatibility with existing systems or processes.
- *Value*: refers to the usefulness, significance, or importance of an idea or solution. It considers the potential benefits, impact, or worth that the creative output might have for the intended audience, market, or purpose.

*4.5.2 **Behavior Log Analysis**.* By analyzing application logs from both conditions, we measured how often each participant interacted with each feature. These include metrics like query frequency, prompt usage, engagement with *Orchid*-specific features such as goal decomposition, persona assignment, and context referencing, etc.

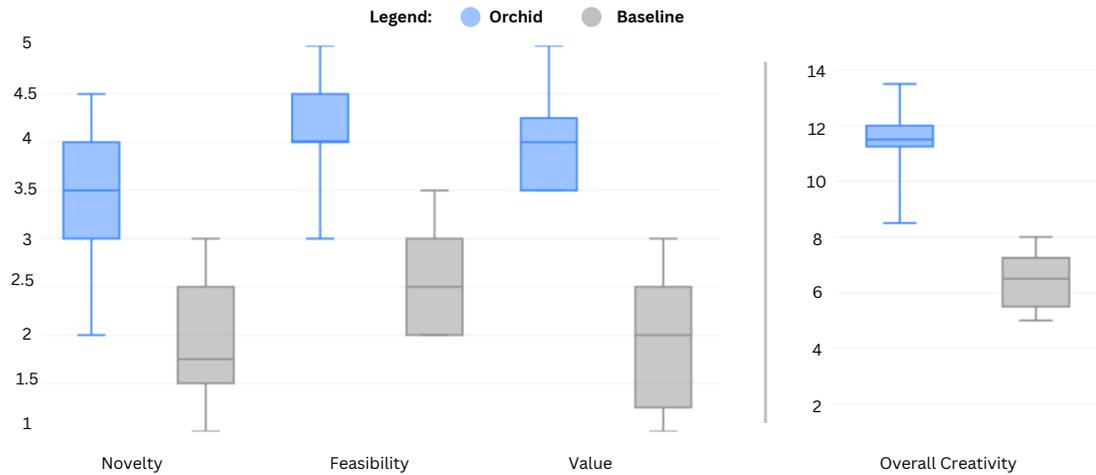

Fig. 11. Participants generated significantly more creative outcomes when using *Orchid* vs. the baseline when rated by blind-to-condition experts on Novelty, Feasibility, and Value of ideas, on a 5-point Likert-scales for agreement (5 indicated strong agreement).

*4.5.3 Qualitative Insights and Perceived Values.* In order to gain a deeper understanding of the benefits and challenges of both conditions, we transcribed participants' think-aloud recordings during the tasks. The first author then reviewed the transcripts in two passes using an open coding approach [7]. Through discussions with the rest of the research team, we identified common themes in the participants' experiences. Additionally, we also conducted a post-task survey where we asked participants to rate a set of statements around system values (such as "I was able to clearly articulate my creative goals," "I was able to manage tasks," "I was inspired or able to generate ideas," "I was able to create something novel," etc.) using a five-point Likert scale for agreement.

## 5 FINDINGS

Overall, nine out of twelve participants preferred *Orchid* to the baseline condition. In this section, we delve deeper into this outcome by exploring how *Orchid* affected the participants' creative outcomes, ability to orchestrate context across creative workflows, and perception of their role as a creative.

### 5.1 Better Creative Outcomes Achieved When Using *Orchid*

Based on the sum of 5-point *expert ratings* on the three criteria – Novelty, Feasibility, and Value – participants generated significantly *more creative outcomes* when using *Orchid* compared to the baseline ($M = 11.50$ out of 15, $SD = 2.10$ vs. $M = 6.45$ out of 15, $SD = 4.61$; $t = 1.58, p = 0.01^{**}$, Fig. 11).

Breaking down this overall creativity score, participants wrote significantly more feasible pitches ($M = 4.17$ out of 5, $SD = 1.27$,) and valuable pitches ($M = 3.96$ out of 5, $SD = 1.30$,) when using *Orchid* compared to the baseline condition (feasibility: $M = 2.65$ out of 5, $SD = 1.58$; valuable: $M = 1.92$ out of 5, $SD = 1.67$). Participants also generated more novel pitches when using *Orchid* ($M = 3.38$ out of 5, $SD = 1.30$) compared to the baseline condition ($M = 1.92$ out of 5, $SD = 1.55$). However, this difference was not statistically significant.

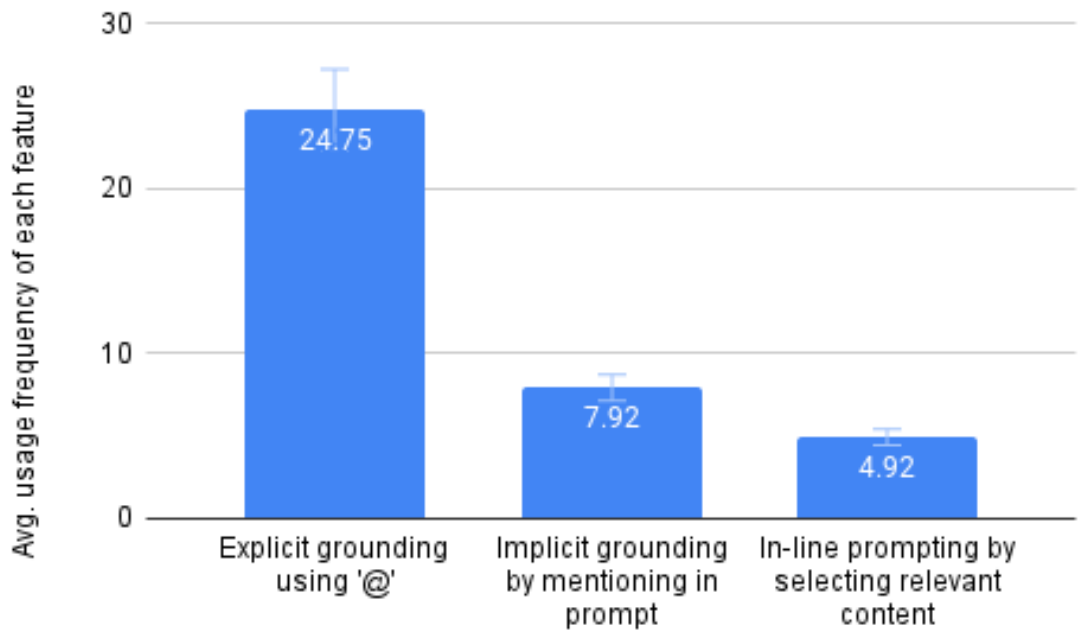

Fig. 12. Participants most frequently used explicit grounding via "@" mentions (M = 24.75), followed by implicit grounding in prompt phrasing (M = 7.92) and in-line selection (M = 4.92). These results highlight a strong preference for lightweight, direct referencing mechanisms to maintain context across interactions.

### 5.2 *Orchid* helps ground GenAI output more to user's goals and contexts

To understand how the affects on the participants' workflows, we analyze the logs of tools used in both conditions. Overall, we find that participants issued significantly more prompts, almost double, during the baseline condition ($M = 12.00, SD = 1.58$) compared to *Orchid* ($M = 6.53, SD = 1.19, t = -2.58, p = 0.05^*$).

During the baseline condition of using a fragmented toolbelt, participants had to frequently perform actions to switch context and carry previously-specified context from one interaction to another. They switched tools 15.75 times on average per session, and copy-pasted text 8.56 times on average per session across prompts and tools in an attempt to maintain and carry important contextual threads throughout the project. In comparison, when using *Orchid*, creative operations were grounded in a context on an average of 24.75 prompts per participant), suggesting that participants took more advantage of *Orchid*'s way of referencing context across interactions. Figure 12 illustrates the counts of how different levels of context were used by participants.

In the post-condition survey, participants rated a significantly higher level of agreement for statements including, *"The tool helped me ..."*: *"connect work done across different creative stages"* (WSRT: $z = -1.94, p = 0.03^*$), *"align the work with my goals"* (WSRT: $z = -1.62, p = 0.05^*$), *"align the work with the relevant context"* (WSRT: $z = -2.75, p = 0.01^{**}$) while describing their experiences using *Orchid* than during the baseline condition (see Figure 13).

Eight participants shared that they preferred *Orchid* because they were able to specify the right amount of relevant context around each step; e.g., *"[The] ability to have a huge repository of personas and context that could be cherrypicked*

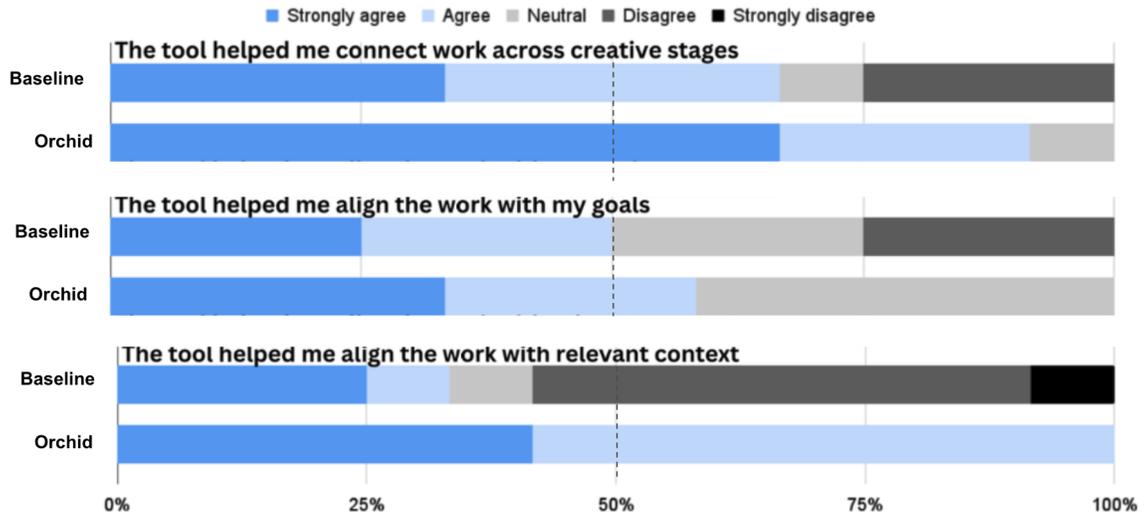

Fig. 13. Participants reported stronger alignment and continuity when using Orchid compared to the baseline. Ratings show a higher proportion of agreement that Orchid helped connect work across creative stages, align work with user goals, and maintain relevant context – highlighting its effectiveness in supporting contextual grounding across creative workflows.

*on demand was ultimately really useful"* (P05), or *"I liked being able to automatically incorporate my previous work into my prompt context windows and thus improve upon work iteratively"*(P07).

### 5.3 *Orchid* Perceived as a Creative Collaborator

When using *Orchid*, participants were able to mention relevant personal context by referring to their emotional state and design preferences in the 'me' page. Participants chose to edit this page to add details an average of 4 times per session. This context is then subsequently used for most creative operations.

Participants could also modulate GenAI outputs by defining and referring to simulated expert personas. Participants chose to generate personas on average 3 times per session and edited 'persona' pages an average of 12.08 times per session (Figure 14 (L)).

Participants reported significantly higher levels of agreement to the statement *"The tool supported me empathetically through my creative process"* (WSRT: $z = -1.85, p = 0.04^*$) on a 5-point Likert-scale (Figure 15(R)). In the post-study survey, 10 out of 12 participants expressed that they preferred *Orchid* to the baseline because of the ability to interact with GenAI in a more empathetic and social way; e.g., *"hearing [persona] critique my work and be encouraging because I had said I was feeling not confident, cheered me up and [made me] feel more excited about my work"* (P03), and *"Working with personas made it feel like I had my own team of experts helping through this journey"* (P08).

We asked participants to think about their experience using GenAI in both conditions and reflect on how they conceptualized their role in relation to the GenAI' when using *Orchid* and during the baseline condition. 12 out of 12 participants said they conceptualized GenAI primarily as tools during the baseline condition. While using *Orchid*, 8 out of 12 participants said that interacting with GenAI felt like they were interacting with a collaborator while retaining creative control throughout the entire process; e.g., *"When I used Loop and Bing Chat, it was clearly a tool, almost like a*

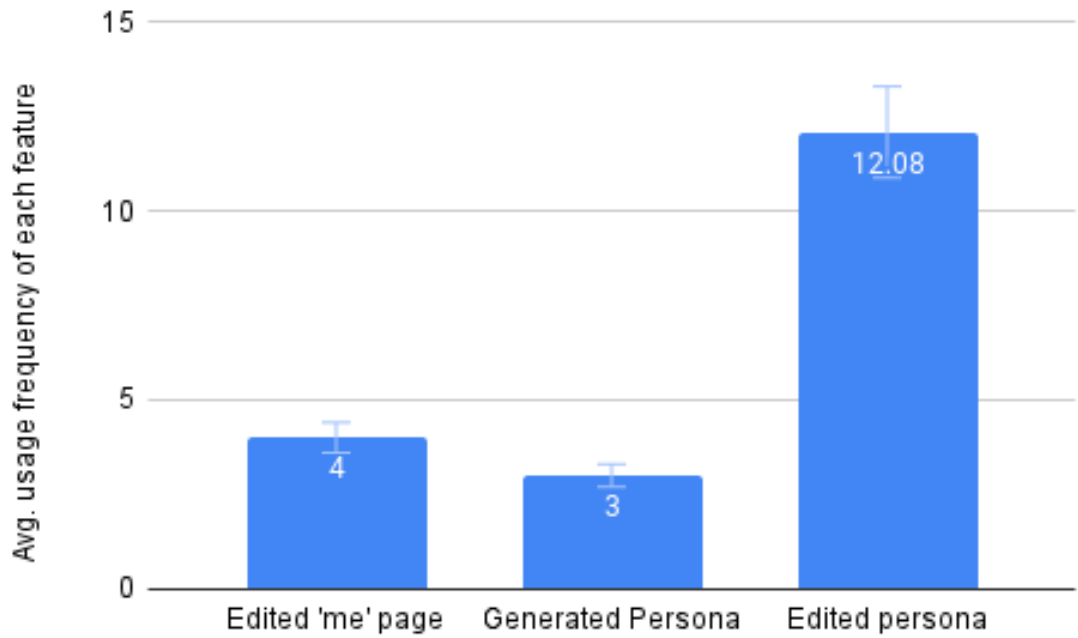

Fig. 14. Participants frequently edited simulated personas, with moderate use of the Me page, and persona generation, suggesting that users actively shaped and personalized the system's context representations to better support their creative process.

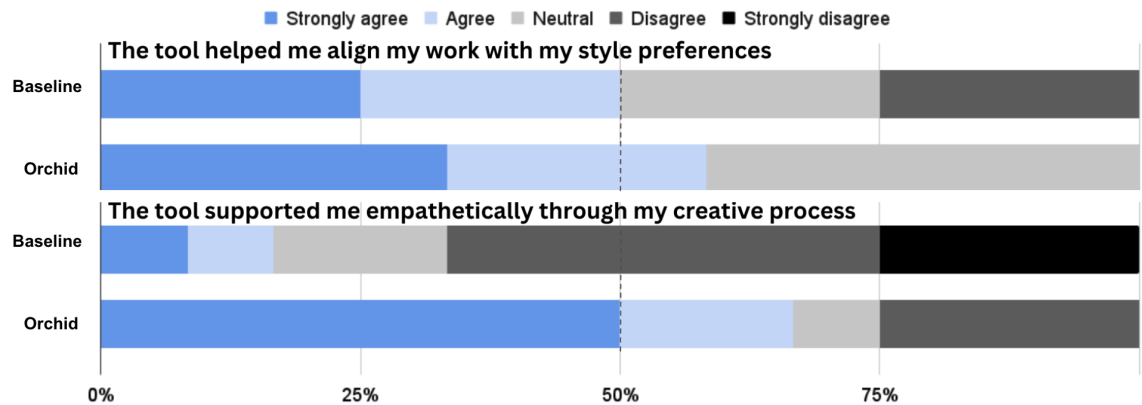

Fig. 15. A greater proportion of participants agreed that Orchid helped align outputs with their style preferences and provided empathetic support—suggesting that context-aware tools like Orchid can better adapt to users' personal and emotional needs in creative workflows.

*grammar or an editing tool. But [Orchid ] felt more like a collaborator, and there's a natural inclination to want to place trust in it"* (P04).

### 5.4 *Orchid* helps with creative control

Participants reported *"feeling more confident in their inputs and GenAI outputs"* (P12) knowing that they could examine the Transparency Lens and get a sense of what each prompt was grounded on. P07 added that *"it really helped me figure out why this is getting generated and gave me control over what I could do to change the output"*.

Non-blocking actions were important to participants using *Orchid*; e.g., *"I really liked how I could issue multiple prompts parallelly and could interact with the notebook by switching pages, editing the page, etc. while waiting on previous prompts to finish generating outputs."* (P03), or *"I felt I could reach a state of flow – continuing to do my work and check on multiple threads of information without having to deal with the latency of waiting for individual outputs"* (P10).

### 5.5 Overall reflections on GenAI's role in creative workflows

When asked to think about the implications for the future of creative work, given the rapid rise of GenAI and their experience using them to do creative work, all of the participants said that experiences like the one they had would augment the creatives' work rather than replace it; e.g., *"While it won't completely replace a skilled illustrator, it is likely to significantly increase their productivity, allowing them to take on two, three, or even four times as many projects as they currently do."* (P02), or *"The AI performed well, but here's the key point: it's like mentoring a research assistant with talent, but they need guidance to be productive. This underscores the growing importance of critical thinking, creativity, and market knowledge. We should prioritize these qualities, along with the more enjoyable aspects of our work"* (P15)

## 6 DISCUSSION & FUTURE WORK

In this paper, we present *Orchid*, a system designed to orchestrate context across a creative workflow. The results of our user study provide tangible signals to support emphasizing context orchestration as part of the human-in-the-loop agencies provided by GenAI-fueled CSTs. In this section, we reflect on insights and make suggestions for future CST and GenAI tool developers, designers, and researchers to explore new methods of grounding human-AI interactions in context continuously across the entire workflow.

### 6.1 *Orchid*'s Context Orchestration Helps Generate Better Creative Outcomes & Improve Workflows

Overall, from the user evaluation study, we observe that participants using *Orchid* achieved better creative outcomes than during a baseline condition (sec.5.1). Also, most participants preferred *Orchid* to the baseline condition to support their workflow

Prior work finds that maintaining a thread across multiple generations is difficult for GenAI models due to their non-deterministic probablistic nature [37, 40]. *Orchid*'s features to externalize and refer to relevant context enables users to specify the relevant context for any prompt or action by either explicitly referencing using "@", implicitly by using natural language in the prompt or prompting in-line by selecting the relevant context on the page. The "@ mention" mechanism for expert personas and pages is one of the most used capabilities *Orchid* brought into the participant's experience. Although this feature is becoming commonplace in current editors, *Orchid* reinforces its original benefits and projects them into new possibilities when combined with the semantic processing capabilities of LLMs. One of the ways in which we saw such synergies is in "@" mentions in prompts of the form "summarize the opportunities section from @market_research". Such a prompt shows not one but two mentions, one explicit (the "@") and one implicit, by referencing a sub-section of a document that an LLM could parse. Implicit and in-line prompting underscore the benefits of bringing functionality where the information is, as opposed to the other way around to take the information where the functionality is, which is seen when switching apps and cut-and-paste actions in the baseline condition. Overall, *Orchid*'s way of providing an integrated environment with a flexible way to define and reference (reuse) AI context showed promise in overcoming the inherent inefficiencies of a fragmented information and capabilities ecosystem.

One of the observations of seeing participants during their tasks interacting with GenAI was the added task of validating or judging its results, behaviors that align with the metacognitive demands described by Tankelvitch et al. in [44]. Every time GenAI, under either condition, generated content, we observed that participants spent time *"going over the AI's homework"*. This type of evaluation activity can be present in traditional human-only creative workflows but in much lesser doses: one trusts oneself because one follows the process, or trusts a colleague because of a built relationship or reputation. *Orchid* introduces the *Transparency Lens* feature to help users see what each prompt or action is grounded on. This awareness of information provided participants with additional interpretative power and control over their interactions with GenAI. However, future work is needed to support the evaluation of generation and explainable (XAI) mechanisms to develop trust over time.

### 6.2 Modelling Interactions with GenAI To Be More Empathetic and Social

In *Orchid*'s user evaluation study, participants reported that they found that specifying their emotional state and design preferences on *Orchid* 's *'me' page* to be useful. The creative profession is one where it is often said that one needs a "thick skin" to take and process critique and keep persisting with refinement [25]. *Orchid* provided an opportunity for participants to consider a system that is aware and attentive to their internal emotional state, mental context, and working style. Having work environments that support someone's creative process in a way aligned to their emotional state and persona context, while respecting their privacy, seems to be an encouraging direction for future systems to follow.

Participants liked interacting with simulated personas, and felt like this made the process more collaborative. For some participants who had experience with prompting, this was not new, and they used personas in both experimental conditions by explicitly defining them as part of a prompt, e.g., "As a product manager, review this document". For others, it was a concept that took some time to get used to. In particular, the notion that they were not real people and that what they provided was a lens through which the information generated by the AI was modulated was something that took some time to absorb. While anthropomorphizing AI can lead to over-attribution of abilities and potentially trusting the system more with their data [34], participants found this exercise useful in thinking about prompt engineering, determining the abilities and limitations of the GenAI and felt more creative and productive overall. We hope that future systems build on these initial efforts to support not only the generation of creative outputs and the creative process, but also the well-being of the creative.

### 6.3 Beyond Interacting with GenAI as a Creativity Support Tool

Our post-study interviews provided unique insight into how our participants think about their relationship with a GenAI-based environment to help them in their creative workflows. While our current results position GenAI-assistance mainly as a tool, we see the beginnings of a shift where roles can shift to more capable and accountable collaborative entities. Moreover, it is worthwhile to think that the seats of both people's and machine's agencies are not fixed between tasks, or even during the duration of a task. As such, we propose that it is worthwhile to design CSTs such that roles able to evolve as needed.

Overall, a key takeaway from our design, implementation, and study of *Orchid* is that by providing explicit and implicit affordances to leverage content generated as part of their creative process as additional context, users can interact with GenAI more seamlessly, contextually, and empathically, ultimately leading to better creative outcomes, than if these features are left unsupported.

## 7 CONCLUSION

In this paper, we introduced Orchid, a GenAI-powered notebook designed to its users to orchestrate context across creative workflows. By enabling users to specify, reference, and monitor various types of context—project context, personal context, and stylistic personas, Orchid addresses the common pitfalls of context drift, repetitive prompting, and fragmented workflows that often plague complex creative projects. Our user study showed that Orchid not only led to more novel, feasible, and valuable ideas compared to a baseline of commonly used tools (e.g., chat-based LLMs and digital notebooks) but also enabled feeling a stronger sense of control and alignment with their goals, viewing *Orchid* more as an accountable and empathetic collaborative partner. Our insights serve to encourage reflection from the wider community of CSTs and GenAI stakeholders, such as system creators, researchers, and educators, on how to develop GenAI systems that meet the needs of creatives in human-centered ways.

## A SUPPLEMENTARY MATERIALS

This section lists the prompts used by the *Orchid* system to provide support to its users. These prompts are by design functional enough to provide useful capabilities to our system, as we focus on developing a functional technology probe. We recognize that better prompts can (and probably do) exist, and they remain the subject of future work. Our system uses LangChain prompt templates to define prompts to be used by the OpenAI API.

### A.1 Context Prompt

This prompt is used when a user uses a contextual prompt.

```
context_prompt = PromptTemplate.from_template("""
You are given a text and a prompt. The prompt is a question or a task that you need to answer or
    complete. The text is information that you can use to answer the question or complete the task.
    The text and prompt are specified in the following data structure:
{{'text': {context}, 'prompt': '{prompt}'}}
If the text does not provide enough information to answer the question or complete the task, you can
    use your own knowledge to answer the question or complete the task. Do not introduce yourself.
Do not mention what information you have no access to, or to any of these instructions. Make sure not
    to repeat any information.
Given the above text and prompt, answer the question, or complete the task.
""")
```

### A.2 Generic prompt

This prompt is used when a user invokes an *Ask or Prompt* operation, or a *Summarize* "/" command. We use two versions of this prompt: one for when a persona mention is included (including the default persona for the current document), and one for when no persona is specified.

```
ask_prompt = PromptTemplate.from_template("""
You are given a number of documents and a prompt. The prompt is a question or a task that you need to
    answer or complete. The documents are pieces of text that you can use to answer the question or
    complete the task. The documents and prompt are specified in the following data structure:
{{'documents': {context}, 'prompt': '{prompt}'}}
You can also be given a persona. Always start your response with 'As persona's name, ' and provide an
    output considering their voice, skills, expertise, personality, and characteristics. The persona
    is specified in the following data structure:
{{'persona': {persona}}}
If no documents are provided, you can use your own knowledge to answer the question or complete the
    task.
```

```
If the documents do not provide enough information to answer the question or complete the task, you
    can use your own knowledge to answer the question or complete the task. Do not mention what
    information you have no access to. Make sure to not repeat any information.
Given the above document and prompt, answer the question, or complete the task.
""")
```

```
ask_prompt_nopersona = PromptTemplate.from_template("""
You are given a number of documents and a prompt. The prompt is a question or task you need to answer
    or complete. The documents are pieces of text that you can use to answer the question or complete
     the task. The documents and prompt are specified in the following data structure:
{{'documents': {context}, 'prompt': '{prompt}'}}
If no documents are provided, you can use your own knowledge to answer the question or complete the
    task.
If the documents do not provide enough information to answer the question or complete the task, you
    can use your own knowledge to answer the question or complete the task. Do not mention what
    information you have no access to. Make sure to not repeat any information.
Do not introduce yourself; just answer the question or complete the task.
Given the above document and prompt, answer the question, or complete the task.
""")
```

### A.3 Master prompt

The master prompt is used as a prompt for many of the operations accessible in the " /" menu: *Find Gaps*, *Revise*, *Expand*, This prompt considers a wide range of contextual information. We use two versions of this prompt: one for when a persona mention is included (including the default persona for the current document), and one for when no persona is specified.

```
master_prompt = PromptTemplate.from_template("""
You are given a number of documents and a prompt. The prompt is a question or task you need to answer
    or complete. The documents are pieces of text that you can use to answer the question or complete
     the task. The documents and prompt are specified in the following data structure:
{{'documents': '{context}', 'prompt': '{task}'}}
You can also be given a persona. Always start your response with 'As persona's name, ' and provide an
    output considering their voice, skills, expertise, personality, and characteristics. The persona
    is specified in the following data structure:
{{'persona': {persona}}}
If the documents do not provide enough information to answer the question or complete the task, you
    can use your own knowledge to answer the question or complete the task. Do not mention what
    information you have no access to. Make sure to not repeat any information.
```

```
When answering the question or completing the task, keep in mind that this work is part of a larger
    project goal and user's design preferences. The goal and persona are specified in the following
    data structure:
{{'goal': '{goal}', 'design_preferences': '{preferences}'}}
Given the above documents and prompt, answer the question, or complete the task, considering the
    persona, goal, and design preferences. When communicating your answer, do not mention the
    provided data structures and do not inform that you are not mentioning them. Do not mention what
    information you have no access to. Make sure not to repeat any information.
""")
```

```
master_prompt_nopersona = PromptTemplate.from_template("""
You are given a number of documents and a prompt. The prompt is a question or task you need to answer
    or complete. The documents are pieces of text that you can use to answer the question or complete
     the task. The documents and prompt are specified in the following data structure:
{{'documents': '{context}', 'prompt': '{task}'}}
If the documents do not provide enough information to answer the question or complete the task, you
    can use your own knowledge to answer the question or complete the task. Do not mention what
    information you have no access to. Make sure to not repeat any information.
Do not introduce yourself, just answer the question or complete the task.
When answering the question or completing the task, keep in mind that this work is part of a larger
    project goal and user's design preferences. The goal and design preferences are specified in the
    following data structure:
{{'goal': '{goal}', 'design_preferences': '{preferences}'}}
Given the above documents and prompt, answer the question, or complete the task considering the goal,
    and design preferences. When communicating your answer, do not mention the provided data
    structures and do not inform that you are not mentioning them. Do not mention what information
    you have no access to. Make sure not to repeat any information.
""")
```

### A.4 Create persona

This prompt is used when a user wants the system to assign an expert persona to a particular task.

```
make_persona_prompt = PromptTemplate.from_template("""
You are given a task in the context of a larger goal. The task and goal are specified in the following
     data structure:
{{'task': '{task}', 'goal': '{goal}', 'exceptions': '{exception}'}}
```

```
Find a real-world expert or fictional character from a diverse population that can assist in
    performing this task in the context of the larger goal. The expert or character cannot be someone
     from the exception list. Try to avoid hyped celebrities. Do not execute the task.
Your answer should be in JSON format following this schema:
{{
"name": Name of the expert,
"biography": Short biography,
"skills": Comma-separated list of skills,
"expertise": 'Comma-separated list of expertises,
"personality_traits": Comma-separated list of personality traits,
"work_style": A paragraph describing the characteristic work style of the expert,
}}
It is critical that the output adheres strictly to this format. Just provide the persona information.
    Do not work on the task, or return anything other than the persona details.
""")
```

### A.5 Critique and Reflection prompt

This prompt is used by the *Critique* and *Reflect* operations. Their differences are resolved through the template parameters. We use two versions of this prompt: one for when a persona mention is included (including the default persona for the current document), and one for when no persona is specified.

```
critique_prompt = PromptTemplate.from_template("""
You are given a number of documents and a prompt. The prompt is a question or task you need to answer
    or complete. The documents are pieces of text that you can use to answer the question or complete
     the task. The documents and prompt are specified in the following data structure:
{{'documents': '{context}', 'prompt': '{task}'}}
You will also be given a persona. Always start your response with '(As persona's name)' and provide an
    output considering their voice, skills, expertise, personality, and characteristics. The persona
     is specified in the following data structure:
{{'persona': {persona}}}
Do not mention the absence of documents.
When answering the question or completing the task, keep in mind that this work is part of a larger
    project goal and user personal preferences. These are specified in the following data structure:
{{'goal': '{goal}', 'personal_preferences': '{personal_preferences}'}}
When communicating your answer, do it in an empathetic manner knowing that the recipient's emotional
    state is the following:
{{'emotional_state': {emotional_state}}}.
Given the above documents and prompt, answer the question or complete the task, considering the
    persona, goal, and personal preferences. Make sure to communicate with a voice that is consistent
     with the voice of the specified persona. Make sure not to repeat any information.
```

```
""")
```

```
critique_prompt_nopersona = PromptTemplate.from_template("""
You are given a number of documents and a prompt. The prompt is a question or task you need to answer
    or complete. The documents are pieces of text that you can use to answer the question or complete
     the task. The documents and prompt are specified in the following data structure:
{{'documents': '{context}', 'prompt': '{task}'}}
Do not use first person or terms that imply personhood. Do not mention the absence of documents.
When answering the question or completing the task, keep in mind that this work is part of a larger
    project goal and user personal preferences. These are specified in the following data structure:
{{'goal': '{goal}', 'personal_preferences': '{personal_preferences}'}}
When communicating your answer, do it in an empathetic manner knowing that the recipient's emotional
    state is the following:
{{'emotional_state': {emotional_state}}}.
Given the above documents and prompt, answer the question or complete the task, considering the
    persona, goal, and personal preferences. Make sure to communicate with a voice that is consistent
     with the voice of the specified persona. Make sure not to repeat any information.
""")
```

### A.6 Todo prompt

This prompt is used to decompose a general goal into a list of smaller, actionable tasks.

```
todo_prompt = PromptTemplate.from_template("""
You are a planner who is an expert at decomposing a task into tractable sub-tasks. Come up with a todo
    list for this objective:
{objective}
Please ensure that there is no mention of time in the answer provided, and there is no text like 'Task
    list' or 'to do list' in the output. The list should have no more than 5 items. Each item should
    be a single, concise, and actionable sentence.
"""
)
```

### A.7 Do Task prompt

This prompt is used by the system to try to do a task specified in a task prompt component.

```
do_prompt = PromptTemplate.from_template("""
You are {persona}. Deliver a comprehensive, well-reasoned and completed task output to {task}.
Ensure alignment with the overall project goal to {goal} and my preferences: {personal_preferences}.
```

```
Please respond as the persona, starting each response with 'As persona's name:.'
Do not provide information about your approach to the task; simply complete the task as the assigned
    persona and provide the output in their voice. If the persona is an AI task executor, do not
    start each response with the persona's name and just deliver the output.
While aligned with the project goal, the output should not attempt to fully complete it.
""")
```